\documentclass[aps,prl,twocolumn,groupedaddress]{revtex4}

\usepackage{graphicx} 
\usepackage{dcolumn}
\usepackage{bm}

\begin{document}
\title{Large oxygen-isotope shift above the quantum critical point of Y$_{1-x}$Ca$_{x}$Ba$_{2}$Cu$_{3}$O$_{7-\delta}$} 
\author{John Mann, Pieder Beeli, and Guo-meng Zhao$^{*}$} 
\affiliation{Department of Physics and Astronomy, 
California State University, Los Angeles, CA 90032, USA}

\begin{abstract}
We have studied the oxygen-isotope effect on the superconducting transition temperature
$T_{c}$ in overdoped Y$_{1-x}$Ca$_{x}$Ba$_{2}$Cu$_{3}$O$_{7-\delta}$ with $x$ 
= 0.10, 0.20, and 0.25. We find the oxygen-isotope exponent $\alpha_{O}$
to be small ($\sim$0.02) for $x$ = 0.10 
but substantial ($\sim$0.1) for $x$ = 0.20 and 0.25. The doping level above which $\alpha_{O}$ increases sharply 
coincides
with a quantum critical point 
where the normal-state pseudogap starts to diminish.  The present
isotope-effect experiments provide direct and quantitative constraints on the pairing mechanism of 
high-temperature superconductivity in cuprates.
 
\end{abstract}
\maketitle 

The importance of phonons in the pairing mechanism can usually be
checked by the dependence of $T_{c}$ on the isotope-mass $M$ (isotope
effect). For most conventional phonon-mediated superconductors, the
isotope exponent ($\alpha$ = $-d\ln T_{c}/d\ln M$) is close to 0.5, as
expected from the BCS theory. For high-$T_{c}$ cuprate superconductors, 
the isotope effects have been investigated by several groups with
consistent results \cite{Batlogg,Bourne87,Don88,Crawford90,Bornemann92,Franck93,ZhaoCu}. There are small oxygen- and
copper-isotope exponents in optimally-doped cuprates, but large
exponents (even larger than 0.5) in underdoped cuprates \cite{ZhaoJPCM}. Due to the 
high $T_c$ values and the earlier observation of a very small oxygen-isotope 
exponent ($\alpha_{O}$ $\simeq$ 0.02) 
in an optimally-doped cuprate superconductor YBa$_{2}$Cu$_{3}$O$_{7-y}$ 
(YBCO)
\cite{Batlogg,Bourne87,Don88}, most researchers 
believe that electron-phonon coupling cannot be the origin of 
90 K superconductivity in YBCO. On the other hand, it is extremely difficult to
understand the large $\alpha_{O}$ 
and the substantial oxygen-isotope effect on the in-plane supercarrier mass
($m^{*}_{ab}$) in underdoped cuprates
\cite{ZhaoJPCM,ZhaoLSCO,ZhaoNature} if the electron-phonon
interactions were not strong. The large oxygen-isotope effects on both
$T_{c}$ and $m^{*}_{ab}$ in underdoped cuprates can be 
consistently explained if one considers that $T_{c}$ in
underdoped cuprates is essentially proportional to
$n_{s}/m^{*}_{ab}$ (where $n_{s}$ is the supercarrier density)
\cite{Uemura} due to a Bose-Einstein-like superconducting transition and 
that $m^{*}_{ab}$ depends on the isotope mass due to a polaronic effect
\cite{Alex}. For optimally-doped YBCO where the superconducting transition 
is mean-field-like, the small $\alpha_{O}$ and
substantial oxygen-isotope effect on $m^{*}_{ab}$ \cite{ZhaoYBCO,Zhaoisotope,Keller1} 
can be consistently explained by a scenario where polarons are bound into the Cooper pairs
\cite{Zhaoisotope}.

Now a question arises: What happens to $\alpha_{O}$ when the polaronic
effect disappears? Electronic specific-heat data indicate that
there exists a
quantum critical point at a doping level $p_{cr}$ of about 0.19 in YBCO and
Ca substituted YBCO, above which
the normal-state pseudogap disappears \cite{Loram}. For
La$_{2-x}$Sr$_{x}$CuO$_{4}$ (LSCO), the normal-state pseudogap
state is still present up to a doping level of 0.24
(Ref.~\cite{Loram}), implying that $p_{cr}$
$>$ 0.24 in this system. Although the origin of the normal-state pseudogap
is still debated, one plausible explanation is that there
coexist Fermi-liquid-like carriers and polaronic/bipolaronic
oxygen holes  and that the pseudogap is related to the
bipolaron binding energy \cite{MullerJPCM}.  Above the quantum
critical point, the pseudogap 
diminishes along with the disappearance of polaronic and bipolaronic oxygen-holes so that the 
superconducting transition becomes more conventional. If the
superconducting pairing is mainly mediated by phonons, the isotope
effect on $T_{c}$ should become large above the quantum critical point.
Therefore, above this critical
point, the recovery of the large conventional isotope exponent would
unambigously expose electron-phonon coupling as an important  pairing mechanism.

Here we report studies of the oxygen-isotope effect on the superconducting transition temperature
$T_{c}$ in overdoped Y$_{1-x}$Ca$_{x}$Ba$_{2}$Cu$_{3}$O$_{7-\delta}$ with $x$ 
= 0.10, 0.20, and 0.25. We find that the oxygen-isotope exponent $\alpha_{O}$
is small
($\sim$0.024) for $x$ = 0.10 and
becomes substantial ($\sim$0.1) for $x$ = 0.20 and 0.25. The doping level above which $\alpha_{O}$ increases sharply appears to coincide
with the quantum critical point 
where the normal-state pseudogap starts to diminish.  We further show 
that a combined mechanism based on strong coupling to multiple phonon modes 
and significant coupling to high-energy
(1.6-2.0 eV) charge fluctuations is in quantitative agreement with the
oxygen-isotope exponent and magnetic penetration depth above the quantum critical point
as well as with thermal-difference reflectance data.

Samples of Y$_{1-x}$Ca$_{x}$Ba$_{2}$Cu$_{3}$O$_{7-\delta}$ were 
prepared by 
a conventional solid-state reaction using CaCO$_{3}$ (99.99$\%$), BaCO$_{3}$
(99.997$\%$), CuO (99.995
$\%$), and Y$_{2}$O$_{3}$ (99.999
$\%$). The 
powders were mixed, ground thoroughly, and 
fired in air at 
950 $^{\circ}$C for $\sim$24 hours with one intermediate grinding.
The samples were ground again, pressed into pellets, and sintered in
air at 920 $^{\circ}$C for $\sim$20 hours. To obtain samples with small grains and sufficient 
porosity for the isotope experiments, we reground the samples 
thoroughly, pressed them into pellets, and annealed them in air at 
800 
$^{\circ}$C for 20 hours.

Three pairs of the samples with different Ca concentrations were wrapped in gold foil and subjected to 
$^{16}$O and $^{18}$O isotope diffusion. The diffusion was carried 
out for about 48 hours 
at 750 $^{\circ}$C and in an oxygen partial pressure of about 0.8 bar with one
intermediate refilling of the isotope gases. The cooling rate 
from 750 $^{\circ}$C to 670 $^{\circ}$C is about 1.7
$^{\circ}$C/minute, and the average cooling rate from 670 $^{\circ}$C 
to 350 $^{\circ}$C is about 10 $^{\circ}$C/minute. The oxygen isotope-enrichment was determined 
from the weight change of the $^{18}$O sample. The 
$^{18}$O samples were thus found to have $\sim$ 96$\%$ $^{18}$O
and $\sim$ 4$\%$ $^{16}$O. 

\begin{figure}[htb]
    \includegraphics[height=5cm]{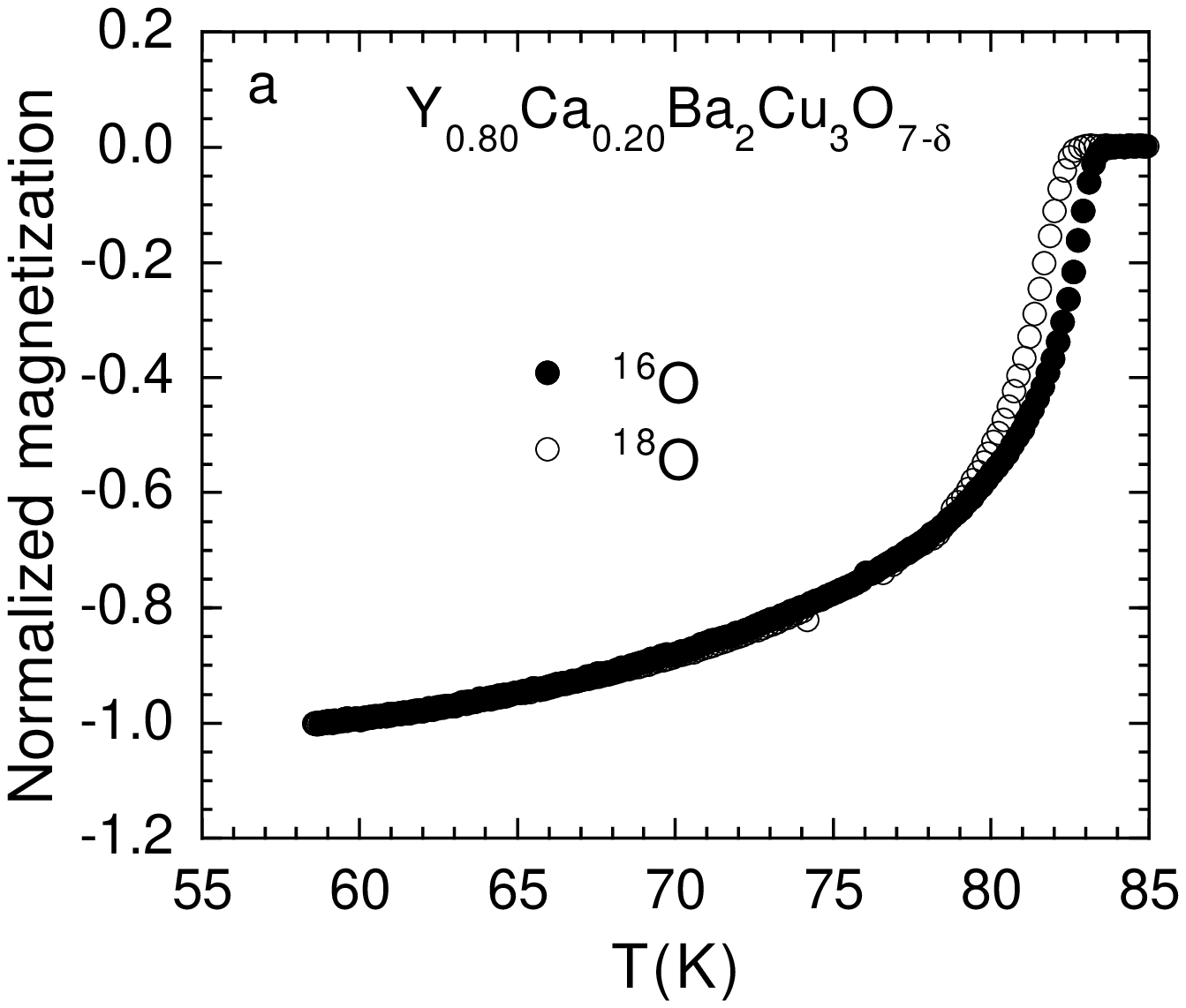}
	 \includegraphics[height=4.7cm]{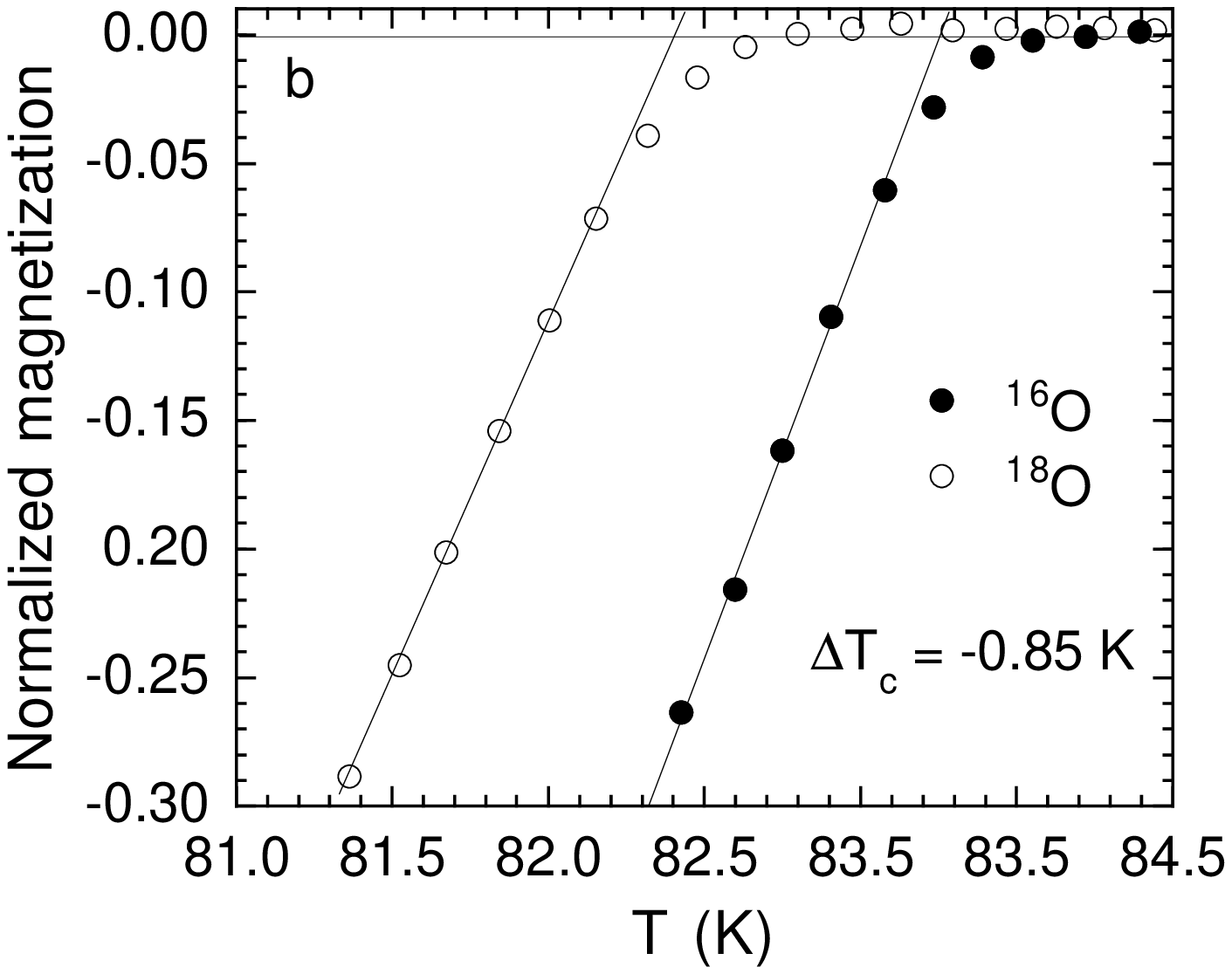}
 \caption[~]{a) Temperature dependencies of the normalized magnetizations 
for the $^{16}$O and $^{18}$O samples of Y$_{0.80}$Ca$_{0.20}$Ba$_{2}$Cu$_{3}$O$_{7-\delta}$.
b) The expanded view near the transition region. The oxygen-isotope
shift is 0.85$\pm$0.04 K.}
\end{figure}

Magnetization was measured with a Quantum Design sample vibrating magnetometer
(VSM). The
Meissner effect was measured in a magnetic field of 1 Oe or 10 Oe 
after the samples had been cooled from the normal state in the field. 
Data were taken continuously during cooling and warming with a same 
rate. For most
measurements, the cooling
and warming rates are 1 K per minute. Such slow cooling and
warming rates ensure a very small thermal lag, which is about 0.5 K.
Increasing the rate to 3 K per minute increases the thermal 
lag to about 1.5 K, but the isotope shift is nearly
independent of the rate.
The data in Fig.~1 and Fig.~2 below are corrected for these thermal
lags.

Figure~1a show temperature dependencies of the normalized magnetizations 
for the $^{16}$O and $^{18}$O samples of Y$_{0.80}$Ca$_{0.20}$Ba$_{2}$Cu$_{3}$O$_{7-\delta}$.
It is apparent that the magnetizations of the two isotope samples show
a parallel shift. In order to see the isotope shift more clearly, we
show in Fig. 1b the expanded view near the transition region. We 
clearly see that the $T_{c}$ of the $^{18}$O 
sample is significantly lower 
than that of the $^{16}$O 
sample. If the transition
temperature is determined from the linear portion of
the magnetization data extended to the base line as indicated
in the figure, we find that the isotope shift is 0.85 K. The
oxygen-isotope exponent $\alpha_{O}$ = $-d\ln T_{c}/d\ln M_{O}$ is calculated to be
0.092, which is larger than the Sn isotope exponent (0.08) in the conventional
phonon-mediated superconductor Nb$_{3}$Sn. In contrast, the isotope shift in Y$_{0.90}$Ca$_{0.10}$Ba$_{2}$Cu$_{3}$O$_{7-\delta}$
is significantly smaller, as seen in Fig.~2. The oxygen-isotope exponent is calculated to be
0.024. The magnitude of the isotope exponent for Y$_{0.90}$Ca$_{0.10}$Ba$_{2}$Cu$_{3}$O$_{7-\delta}$ is very 
close to that (0.027) for optimally doped YBa$_{2}$Cu$_{3}$O$_{7-\delta}$
(Ref.~\cite{ZhaoYBCO}). 

\begin{figure}[htb]
    \includegraphics[height=5cm]{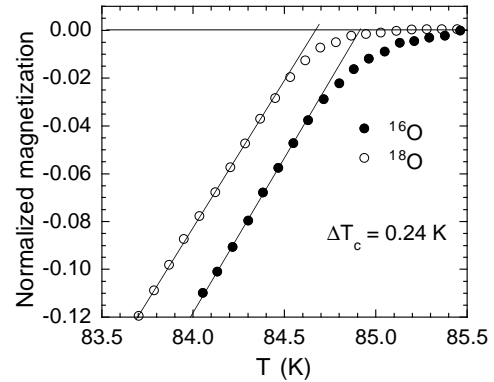}
 \caption[~]{a) Temperature dependencies of the normalized magnetizations 
for the $^{16}$O and $^{18}$O samples of Y$_{0.90}$Ca$_{0.10}$Ba$_{2}$Cu$_{3}$O$_{7-\delta}$.
The oxygen-isotope
shift is 0.24$\pm$0.04 K.}
\end{figure}

In Figure 3, we plot $\alpha_{O}$ and $T_{c}$ as a function of Ca content
$x$. The data for $x$ = 0 are taken from
Ref.~\cite{ZhaoYBCO}. It is striking that the isotope exponent remains
constant and small for $x$ between 0 and 0.10, and jumps to large 
values (about 0.1) for $x$ = 0.20 and 0.25. We will show that this
unusual doping dependence is associated with a quantum critical point 
at a critical doping level of about 0.19.

\begin{figure}[htb]
    \includegraphics[height=5cm]{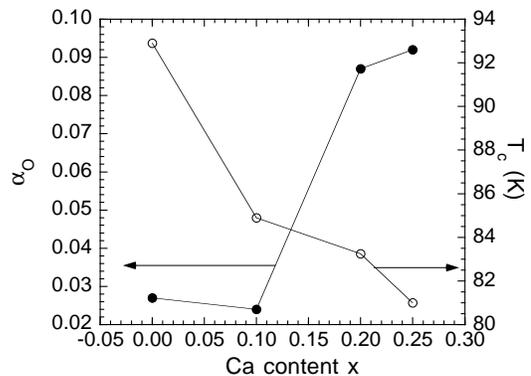}
 \caption[~]{The oxygen-isotope
exponent $\alpha_{O}$ and $T_{c}$ as a function of Ca content
$x$ in Y$_{1-x}$Ca$_{x}$Ba$_{2}$Cu$_{3}$O$_{7-\delta}$. The data for $x$ = 0 are taken from
Ref.~\cite{ZhaoYBCO}.}
\end{figure}

The oxygen content of our samples can be estimated from the final cooling
rate we use to prepare the samples. Studies of the effect of the cooling rate on
the oxygen content in ceramic samples of YBa$_{2}$Cu$_{3}$O$_{7-\delta}$
indicate that the oxygen content at a cooling rate of 10 $^{\circ}$C/minute 
has no differnce from that at a cooling rate of 1 $^{\circ}$C/minute. It is 
also established \cite{Jorgensen} that the $\delta$ value is about 0.07-0.08
for ceramic samples cooled in 1 bar oxygen partial pressure and at a cooling rate of about 1 $^{\circ}$C/minute.
Therefore, the $\delta$ value of YBa$_{2}$Cu$_{3}$O$_{7-\delta}$
should be about 0.1 when it is cooled in 0.8 bar oxygen partial pressure and
at the cooling rate of 10 $^{\circ}$C/minute. It is also shown
\cite{Fisher} that Ca substitution 
does not change the average Cu valence so that $\delta =
\delta_{0} + x/2$, where $\delta_{0}$ is the concentration of the oxygen
vacancies for $x$ = 0. The bond-valence sum calculations (see Fig.~4
of Ref.~\cite{Tallon}) indicate that the oxygen vacancies do not
equally reduce the hole doping levels of CuO$_{2}$ planes and CuO
chains while Ca substitution contributes holes mainly to CuO$_{2}$ planes.
Therefore, Ca substitution leads to overdoping of CuO$_{2}$ planes and
underdoping of CuO chains to keep the same average Cu valence. With $\delta_{0}$
= 0.1 estimated from our preparation conditions,  we find the $\delta$ values of our Ca
substituted samples to be 0.15, 0.20, and 0.225 for $x$ = 0.10,
0.20, and 0.25, respectively.

Figure 4 shows the $\delta$ values of our Ca substituted samples as a 
function of $x$ together with the $\delta_{OP}$ values corresponding to the 
optimal $T_{c}$'s with $p$ = 0.16. It is apparent that the difference 
between $\delta_{OP}$ and $\delta$ becomes more pronounced above $x$ =
0.13. We can estimate the hole doping level $p$ of CuO$_{2}$ planes
from the bond-valence sum calculations which show that $dp/d\delta$ = 
$-$0.20 
for $\delta$ $<$ 0.5. Using $dp/d\delta$ = $-$0.20 and the
differnce $\delta_{OP} - \delta$ (see Fig.~4), we 
calculate the $p$ values to be 0.18 and 0.21
for $x$ = 0.10 and 0.20, respectively. With $p_{cr}$ = 0.19 for
the double-layer cuprates, one can see that the $x$ = 0.10 sample
is  below the quantum critical point and the $x$ = 0.20 sample is 
above it.
\begin{figure}[htb]
    \includegraphics[height=5cm]{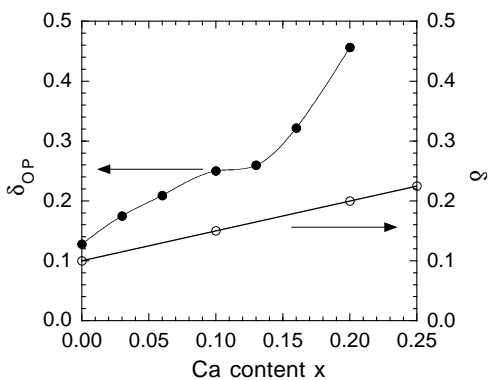}
 \caption[~]{The concentration $\delta$ of oxygen vacancies in Y$_{1-x}$Ca$_{x}$Ba$_{2}$Cu$_{3}$O$_{7-\delta}$ as a 
function of $x$ together with the $\delta_{OP}$ values corresonding to the 
optimal $T_{c}$'s with $p$ = 0.16. The $\delta_{OP}$ data are from
Ref.~\cite{Tallon}. }
\end{figure}

If the normal-state pseudogap is associated with local pairs of polaronic 
oxygen holes, the sudden increase of the oxygen-isotope exponent just above the quantum critical
point, where the normal-state pseudogap diminishes, is consistent with the disappearance of 
polaronic oxygen holes. Diminishing polaronic effect above $p_{cr}$ makes the
superconducting transition more conventional so that the total isotope
exponent will be close to 0.5 if electron-phonon coupling mainly
contributes to the electron pairing. The fact that the observed oxygen-isotope exponent 
above $p_{cr}$ is substantial (about 0.1), but still significantly below 0.5,
implies that other non-oxygen related phonon modes and/or high-energy electronic
bosonic modes also contribute to the electron pairing. On the assumption
that the electron-phonon spectral function $\alpha^{2} (\omega) F(\omega)$
is simply proportional to the phonon density of states measured 
by neutron inelastic scattering, Rammar \cite{Rammar} calculated the oxygen-isotope exponent 
to be 0.08-0.10 for an electron-phonon coupling
constant $\lambda_{e-ph}$ in the range of 1.4 to 6.5 and effective
Coulomb pseudopential $\mu_{eff}^*$ in the range of $-$0.15 to 0.30.
A negative $\mu_{eff}^*$ value is required to obtain 91 K
superconductivity for smaller $\lambda_{e-ph}$. For example, $\mu_{eff}^*$
= $-$0.15 when $\lambda_{e-ph}$ = 1.4. Within the combined phonon and
nonphonon mechanism, the negative $\mu_{eff}^*$ = $\mu^* -
\lambda_{e-el}$ (Ref.~\cite{Carb}) is possible only when the coupling constant $\lambda_{e-el}$
of high-energy 
electronic bosonic modes is larger than the Coulomb pseudopential
$\mu^*$, which has a typical value of 0.10-0.15 for most conventional 
superconductors. Our
measured $\alpha_{O}$ is in quantitative agreement with the calculated
values (0.08-0.10) that are nearly independent of the $\mu_{eff}^*$ values. 
Furthermore, the simple assumption of the multiple-phonon 
coupling made by Rammar \cite{Rammar} is consistent with
tunneling and angle-resolved photoemission spectra, which show
multiple-phonon coupling
features that precisely match with the peak features in the phonon
density of states \cite{Zhaopair}.   

The relative contribution of electron-phonon coupling in
superconducting pairing
depends on the average electron-phonon coupling constant. The tunneling
spectrum \cite{Gonnelli} of optimally doped Bi$_{2}$Sr$_{2}$CaCu$_{2}$O$_{8+y}$
indicates that
the electron-phonon coupling constant is about 3. Since this
tunneling spectrum may mainly probe the anti-nodal states that have the
maximum coupling constant, the average coupling constant over the
entire Fermi surface should be
approximately 1.5 (the half of the maximum value)  due to extended-$s$ wave gap symmetry with eight line nodes
\cite{Zhaosymmetry,Zhao2007}. For the fully oxygenated YBa$_{2}$Cu$_{3}$O$_{7}$ with
$T_{c}$ = 89 K, the doping level $p$ is just at the
critical point \cite{Loram}, suggesting negligible polaronic effect.
The in-plane penetration depth for this compound \cite{Hardy} is found to be 103$\pm$8 nm,
corresponding to an effective plasma energy of 1.92$\pm$0.14 eV. With 
a bare in-plane plasma energy of 2.9 eV (Ref.~\cite{Allen}), we find that the masses of
charge carriers are enhanced by a factor of 2.28$\pm$0.35, implying 
that the average $\lambda_{e-ph}$ is 1.28$\pm$0.35. With $\lambda_{e-ph}$
= 1.3, it is required to have $\mu_{eff}^*$ = $-$0.16 to obtain 89 K
superconductivity. Indeed, thermal-difference
reflectance data of several cuprates show that \cite{Little2} the coupling constant
to high-energy (1.6-2.0 eV) charge fluctuations is in the range of
0.25-0.30. This implies that $\mu_{eff}^*$ should be in the range of
$-$0.10
to $-$0.20 with $\mu^*$ = 0.10-0.15. Therefore, the strong interactions of electrons with both
phonons and charge fluctuations can cause 90 K superconductivity.

Now we discuss the origin of the quantum critical point. It is
well known that undoped parent compounds are charge-transfer insulators due to
strong electron-electron correlation. Therefore, doped holes mainly reside on the oxygen orbitals as 
long as the charge-transfer gap is larger than a critical value. The doped oxygen holes can 
form polarons and 
bipolarons due to strong electron-phonon interactions \cite{Alex}. For
LSCO, this should happen for $p$ $<$ 0.10 (Ref.~\cite{MullerJPCM}).  When the doping level increases, the 
charge-transfer gap gradually decreases \cite{Raim}. Above a critical 
doping level $p_{cr}$, the charge-transfer gap becomes small enough to lead to 
the formation of a single 
band, similar to that predicted from the calculation of local density approximation. 
In this doping regime, both the polaronic oxygen holes and normal-state
pseudogap 
disappear and a conventional Fermi-liquid state is restored. For LSCO, $p_{cr}$
is larger than 0.24 while  $p_{cr}$ $\simeq$ 0.19 for YBCO~\cite{Loram}. 
In the intermediate region (0.10 $<$ $p$ $<$ $p_{cr}$), 
there coexist polaronic/bipolaronic oxygen holes and Fermi-liquid carriers 
\cite{MullerJPCM} and
the interaction between the two types of carriers leads to
non-Fermi-liquid behavior. The Bose-Einstein condensation of local
oxygen-hole pairs leads to $d$-wave order-parameter (OP) symmetry
\cite{Alex98} in the
underdoped region, consistent with some phase-sensitive experiments
\cite{Tsuei}
which probe a $d$-wave OP symmetry of underdoped surfaces and/or interfaces
\cite{Bet,Mann}.

In summary, we have studied the oxygen-isotope effect on the superconducting transition temperature
$T_{c}$ in overdoped Y$_{1-x}$Ca$_{x}$Ba$_{2}$Cu$_{3}$O$_{7-\delta}$ with $x$ 
= 0.10, 0.20, and 0.25. We find that the oxygen-isotope exponent $\alpha_{O}$
is small ($\sim$0.02) for $x$ = 0.10 and
becomes substantial ($\sim$0.1) for $x$ = 0.20 and 0.25. The doping level above which $\alpha_{O}$ increases sharply 
coincides
with the quantum critical point 
where the normal-state pseudogap starts to diminish.  We further show 
that a combined mechanism based on strong coupling to multiple phonon modes 
and significant coupling to high-energy
(1.6-2.0 eV) charge fluctuations is in quantitative agreement with
the oxygen-isotope exponent and magnetic penetration depth above the 
quantum critical point \cite{Hardy} as well as with thermal-difference reflectance data
\cite{Little2}.
 ~\\
 ~\\
\noindent 
{\bf 
Acknowledgment:} This research is supported by a Cottrell Science Award from Research Corporation. We thank the Palmdale Institute of Technology for the use of the VSM and 
Lockheed Martin Aeronautics for the cryogens.\\
~\\
$^{*}$ gzhao2@calstatela.edu
\bibliographystyle{prsty}

\end{document}